\documentclass[12pt,a4paper]{article}
\usepackage{amsmath,amssymb,type1cm,booktabs,bm,braket}
\usepackage[dvips]{graphicx}
\usepackage{amsmath,amssymb}
\usepackage{comment}	
\graphicspath{{./figures/}}

\allowdisplaybreaks[1]
\numberwithin{equation}{section}
\def\Lim{\qopname\relax\@empty{lim}\limits}
\let\lim = \Lim

\def\sup{\mathop{\operator@font sup}\limits}
\def\inf{\mathop{\operator@font inf}\limits}
\def\max{\mathop{\operator@font max}\limits}
\def\min{\mathop{\operator@font min}\limits}
\def\prod{\mathop{\mathchoice{\textstyle\prod@}{\textstyle\prod@}{%
      \scriptstyle\prod@}{\scriptscriptstyle\prod@}}\limits}
\def\coprod{\mathop{\mathchoice{\textstyle\coprod@}{\textstyle\coprod@}{%
      \scriptstyle\coprod@}{\scriptscriptstyle\coprod@}}\limits}
\def\bigcap{\mathop{\mathchoice{\textstyle\bigcap@}{\textstyle\bigcap@}{%
      \scriptstyle\bigcap@}{\scriptscriptstyle\bigcap@}}\limits}
\def\bigcup{\mathop{\mathchoice{\textstyle\bigcup@}{\textstyle\bigcup@}{%
      \scriptstyle\bigcup@}{\scriptscriptstyle\bigcup@}}\limits}
\def\Int{\displaystyle\intop\ilimits@}
\let\int = \Int
\def\bigoplus{\mathop{\mathchoice{\textstyle\bigoplus@}{%
      \textstyle\bigoplus@}{\scriptstyle\bigoplus@}{%
      \scriptscriptstyle\bigoplus@}}\limits}
\newcommand{\siml}[0]{\hspace{0.3em}\raisebox{0.5ex}
	{$<$}\hspace{-0.8em}\raisebox{-0.75ex}{{\footnotesize 〜}}\hspace{0.4em}}
\newcommand{\simg}[0]{\hspace{0.3em}\raisebox{0.5ex}
	{$>$}\hspace{-0.8em}\raisebox{-0.75ex}{{\footnotesize 〜}}\hspace{0.4em}}

\def\slashchar#1{\setbox0=\hbox{$#1$}	
\dimen0=\wd0				
\setbox1=\hbox{/} \dimen1=\wd1		
\ifdim\dimen0>\dimen1			
\rlap{\hbox to \dimen0{\hfil/\hfil}}	
#1					
\else 					
\rlap{\hbox to \dimen1{\hfil$#1$\hfil}}	
/					
\fi}
\begin{document}
\begin{titlepage}

 \begin{flushright}
 \end{flushright}

 \vspace{1ex}

 \begin{center}

  {\LARGE\bf Baryogenesis by $B-L$ generation due to superheavy particle decay}

  \vspace{3ex}

  {\large $^a$Seishi Enomoto
  \footnote{e-mail: enomoto@eken.phys.nagoya-u.ac.jp},
  $^{ab}$Nobuhiro Maekawa
  \footnote{e-mail: maekawa@eken.phys.nagoya-u.ac.jp}}

  \vspace{4ex}
  {\it$^a$ Department of Physics, Nagoya University, Nagoya 464-8602, Japan}
  \\
  {\it$^b$ Kobayashi Maskawa Institute, Nagoya University, Nagoya 464-8602, Japan}
  \\
  \vspace{6ex}

 \end{center}

 \begin{abstract}
   We have shown that the $B-L$ generation due to the decay of the 
   thermally produced superheavy fields can explain the Baryon assymmetry in the universe
   if the superheavy fields are heavier than $10^{13-14}$ GeV.
   Note that although the superheavy fields have non-vanishing charges under 
   the standard model gauge interactions, the thermally prduced baryon asymmetry is sizable. 
   The $B-L$ violating effective operators induced by integrating the 
   superheavy fields have dimension 7, while the operator in the famous 
   leptogenesis has dimension 5. Therefore, the constraints from the nucleon
   stability can be easily satisfied.
\end{abstract}

\end{titlepage}
\section{Introduction}
To understand the origin of Baryon number $B$ in the universe is one of the most 
interesting subjects in the particle cosmology.
The abundance of the Baryon in the universe is estimated by the nucleosynthesis analysis\cite{nucleosynthesis}
and is observed by the WMAP\cite{WMAP}, and it is quite impressive that they have given the consistent
value for the Baryon density in the universe, which is roughly
\begin{equation}
\frac{n_B}{s}\sim 10^{-10},
\end{equation}
where $n_B$ and $s$ are the Baryon number density and the entropy density, respectively.
After Sakharov\cite{Sakharov} pointed out the three conditions for the generation of 
the Baryon number $B$ in the universe, many mechanisms for baryogenesis
have been studied in the literature\cite{GUT,Lepton,EW,AD}.
One of the most attractive scenario for the baryogenesis is the GUT baryogenesis\cite{GUT}
in which the decay of superheavy gauge bosons and Higgs appeared in the GUT 
produces the baryon number. Unfortunately, the produced baryon number is
known to be washed out by the sphaleron process\cite{sphaleron} in the standard model (SM).
Since the sphaleron process conserves the $B-L$ number, it is 
important
to produce non-vanishing $B-L$ number. The most famous scenario to produce $B-L$ number is 
the leptogenesis\cite{Lepton}, in which the lepton number $L$ is produced by the decay of
the right-handed neutrino. Especially, thermal leptogenesis, in which the
lepton number is produced by the decay of the right-handed neutrino produced 
thermally, is one of the most interesting scenario because the observed baryon
number can be related with the measurements on neutrino masses and mixings.
However, since the scenarios in which the thermal leptogenesis can be applied
are limited, other possibilities to produce the $B-L$ number are worth considering. 
In this paper, we study the $B-L$ production by the decay of certain superheavy
fields with intermediate masses, which can be a remnant of some GUT models.

\section{$B - L$ Violating Interactions}

\begin{figure}[t]
 \begin{center}
  \includegraphics[scale=0.7]{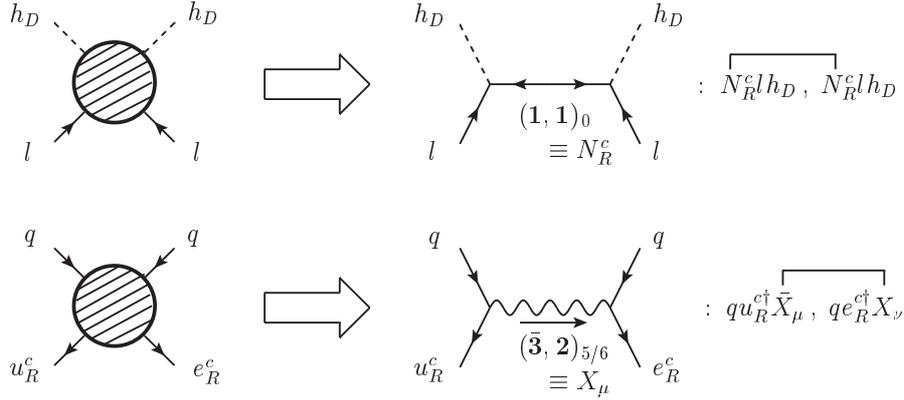}
  \caption{The decomposition of $llh_Dh_D$ (upside) and $qqu_R^ce_R^c$ (downside).  The \textit{fook} over operators means the contraction.}
  \label{fig:llhh_qque}
 \end{center}
\end{figure}

In the SM,
the renormalizable operators cannot break the $B$ and $L$ numbers.
Therefore, the non-conserving
interactions appear in the higher dimensional operators whose mass 
dimensionis larger than four.  For example,
the dimension five operators $llh_Dh_D$ between the doublet lepton $l$ and the 
doublet Higgs $h_D$ have non-vanishing $B-L$ charges and give neutrinos masses.
The dimension 6 operators $qqu^{c\dagger}_Re^{c\dagger}_R$ break 
the $B$ and $L$ numbers, which can induce the proton decay.
(For our notation of the particle contents, see Table \ref{table:contents}.)  
\begin{table}[t]
 \begin{center}
  \begin{tabular}{|l|c||c|c|c|} \hline
   \multicolumn{2}{|c||}{Names} & $( SU(3)_C, SU(2)_L )_{U(1)_Y}$ & $B$ & $L$ \\ \hline \hline
   doublet Quark & $q$ & $( \mathbf{3}, \mathbf{2} )_{1/6}$ & $+1/3$ & $0$ \\
   right-handed Up & $u^c_R$ & $( \bar{\mathbf{3}}, \mathbf{1} )_{-2/3}$ & $-1/3$ & $0$ \\
   right-handed Down & $d^c_R$ & $( \bar{\mathbf{3}}, \mathbf{1} )_{1/3}$ & $-1/3$ & $0$ \\
   doublet Lepton & $l$ & $( \mathbf{1}, \mathbf{2} )_{-1/2}$ & $0$ & $+1$ \\
   right-handed Electron & $e^c_R$ & $( \mathbf{1}, \mathbf{1} )_{1}$ & $0$ & $-1$ \\
   doublet Higgs & $h_D$ & $( \mathbf{1}, \mathbf{2} )_{1/2}$ & $0$ & $0$ \\ \hline
  \end{tabular}  
  \caption{The particle contents and charges.}
  \label{table:contents}
 \end{center}
\end{table}
$llh_Dh_D$ violates $L$ and $B-L$, while  $qqu^{c\dagger}_Re^{c\dagger}_R$ violates $B$ 
and $L$ but not $B - L$.
These higher dimensional operators, $llh_Dh_D$ and $qqu^{c\dagger}_Re^{c\dagger}_R$,
can be induced by integrating the superheavy
right-handed neutrino $\nu_R^c$ and the GUT gauge boson $X$, respectively, as 
in Fig.\ref{fig:llhh_qque}.
The right-handed neutrino plays an important role in the leptogenesis scenario.
And the $X$ gauge boson also plays a crucial role in the GUT baryogenesis.
Therefore, in order to produce $B-L$ number, it must be important to understand which
superheavy particles can induce the $B-L$ violating higher dimensional operators. 

Now, we discuss on other $B - L$ violating operators than $llh_Dh_D$. 
In the literature, in the context of the nucleon decay, $B$ and/or $L$ violating operators have
been classified in the SM\cite{SM} and in the minimal supersymmetric SM (MSSM)\cite{MSSM}.
In the SM, there is no dimension six $B - L$ non-conserving operator.  
It is in dimension seven that we can find out $B - L$ non-conserving operators
\begin{eqnarray*}
&qd_R^cllh_D& \: , \: u_R^cd_R^cd_R^clh_D \: , \: e_R^clllh_D\: ,\:
qqd_R^{c\dagger}l^{\dagger}h_D^{\dagger} \: , \: qu_R^cl^{\dagger}l^{\dagger}h_D^{\dagger} \: , \:
qe_R^cd_R^{c\dagger}d_R^{c\dagger}h_D^{\dagger} \: , \\
&u_R^ce_R^cd_R^{c\dagger}l^{\dagger}h_D^{\dagger}& \: , \: d_R^cd_R^cd_R^clh_D^{\dagger} \: ,\:
qd_R^{c\dagger}d_R^{c\dagger}l^{\dagger} \: , \: u_R^cd_R^{c\dagger}l^{\dagger}l^{\dagger} \: , \:
e_R^cd_R^{c\dagger}d_R^{c\dagger}d_R^{c\dagger} \: .
\end{eqnarray*}
The last three operators include a differential operator or a gauge field. Since the differential
operator can be replaced by the light fermion mass by using the equation of motion,
the contribution of these operators become negligible and we do not consider 
the last three operators in the followings.

\begin{figure}[t]
 \begin{center}
  \includegraphics[scale=0.7]{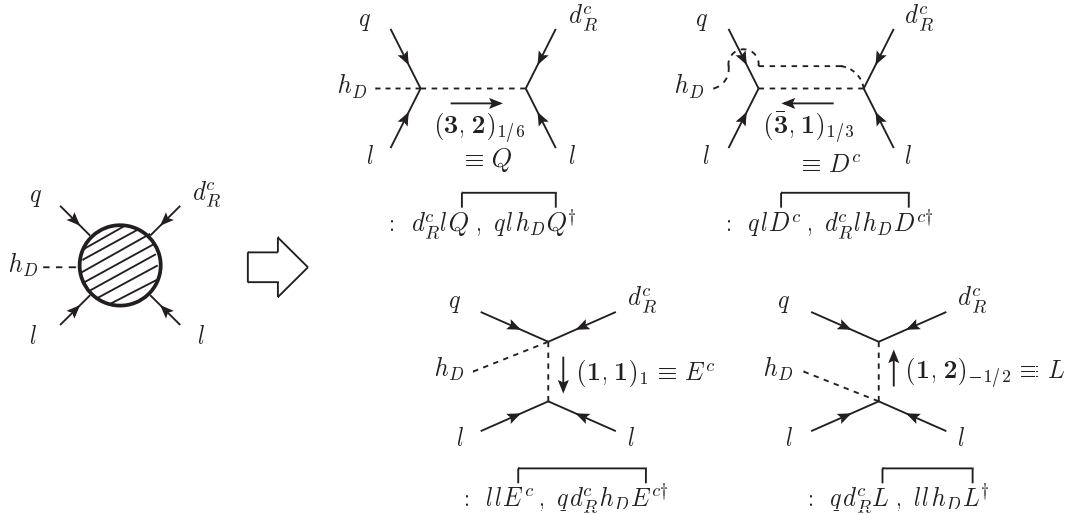}
  \caption{The decompositons of $qd_R^cllh_D$.}
  \label{fig:qdllh}
 \end{center}
\end{figure}

\begin{figure}[h]
 \begin{center}
  \includegraphics[scale=0.7]{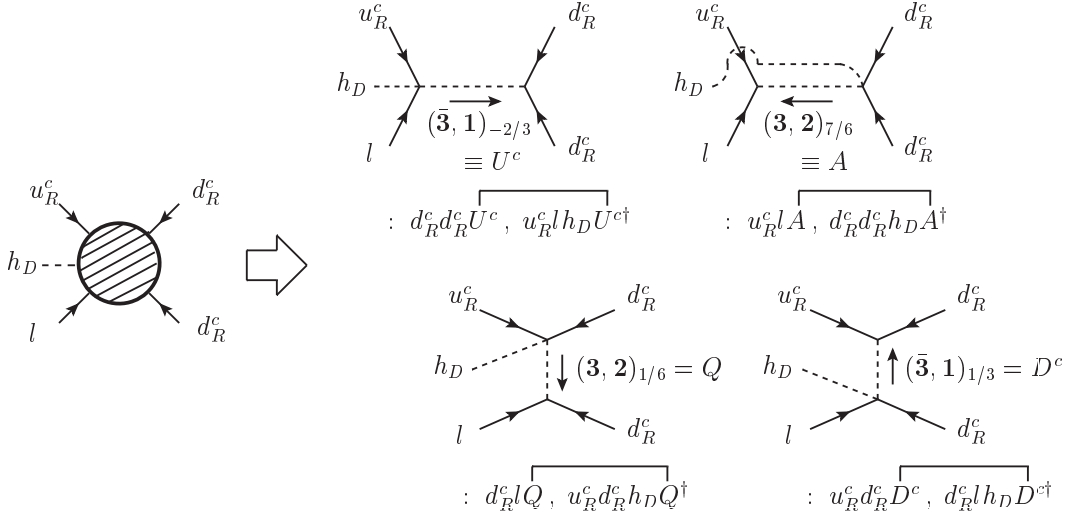}
  \caption{The decompositions of $u_R^cd_R^cd_R^clh_D$.}
  \label{fig:uddlh}
 \end{center}
\end{figure}

Which particles can induce these $B-L$ violating higher dimensional operators?
To answer this question, let us decompose these operators into two parts.
It is useful to write down these operators with
the $SU(5)$ complete multiplets, ${\bf 10}\equiv (q, u_R^c, e_R^c)$, 
${\bf \bar 5}\equiv (d_R^c, l)$, and ${\bf 5}_s\equiv (H_T, h_D)$, where $H_T$ is a colored Higgs,
as ${\bf 10}\cdot{\bf \bar 5}\cdot {\bf \bar 5}\cdot{\bf\bar 5}\cdot{\bf 5}_s$,
${\bf 10}\cdot{\bf 10}\cdot {\bf \bar 5}^\dagger\cdot{\bf\bar 5}^\dagger\cdot{\bf 5}^\dagger_s$, and ${\bf \bar 5}\cdot{\bf \bar 5}\cdot {\bf \bar 5}\cdot{\bf\bar 5}\cdot{\bf 5}_s^\dagger$.
First of all, supposing that the superheavy fields are scalar. Then each part must
includes two fermions. Therefore, the decomposition is limited. For example, 
the operator 
${\bf 10}\cdot{\bf \bar 5}\cdot {\bf \bar 5}\cdot{\bf\bar 5}\cdot{\bf 5}_s$
can be docomposed as $[{\bf 10}\cdot{\bf\bar 5}\cdot{\bf 5}_s+{\bf\bar 5}\cdot{\bf\bar 5}]$ or 
$[{\bf 10}\cdot{\bf\bar 5}+ {\bf\bar 5}\cdot{\bf\bar 5}\cdot{\bf 5}_s]$.
For simplicity, we assume that the superheavy fields are included in the $SU(5)$ multiplets,
${\bf 1}$, ${\bf 24}$, ${\bf 10}$, ${\bf  5}$.
(Though it is straightforward to extend the superheavy fields with the general 
representations, we do not discuss the extension in detail in this paper.)
For the former decomposition, the superheavy scalar belongs to ${\bf 10}$ representation of $SU(5)$, and for the latter, ${\bf  5}$. The concrete decompositions of the operators, $qd_R^cllh_D$ and
$u_R^cd_R^cd_R^clh_D$, can be seen  in Fig.\ref{fig:qdllh} and \ref{fig:uddlh}, respectively.
We denote the superheavy fields as the large characters of the SM
fields which have the same quantum numbers under the SM gauge interactions.
(In Fig.\ref{fig:uddlh}, the superheavy field $A$ has charges of $(\mathbf{3},\mathbf{2})_{7/6}$
under $SU(3)_C\times SU(2)_L\times U(1)_Y$, which belongs to $\mathbf{45}$
of $SU(5)$.)
It is obvious that the superheavy scalars whose representations are ${\bf 10}$ or ${\bf 5}$ also induce 
the other $B-L$ violating operators.
Therefore, we can consider the $B-L$ generation by the
decay of the superheavy scalar fields which belong to ${\bf 10}$ and/or ${\bf 5}$
of $SU(5)$. We will return to this scenario in the next section.

If the superheavy fields are fermions, these operators must be decomposed as
three fermions and one fermion. For example, the operator
${\bf 10}\cdot{\bf \bar 5}\cdot {\bf \bar 5}\cdot{\bf\bar 5}\cdot{\bf 5}_s$
can be decomposed as 
$[{\bf 10}\cdot{\bf \bar 5}\cdot {\bf \bar 5}+{\bf\bar 5}\cdot{\bf 5}_s]$ or
$[{\bf \bar 5}\cdot {\bf \bar 5}\cdot{\bf\bar 5}+{\bf 10}\cdot{\bf 5}_s]$.
For the former decomposition, the superheavy fermions belong to ${\bf 1}$
or ${\bf 24}$, and for the latter, they belongs to ${\bf 10}$ and the complex
conjugate. Actually the right-handed neutrinos which belongs to ${\bf 1}$ of 
$SU(5)$ can induce some of these operators. 
 Though it must be possible to induce non-vanishing 
$B-L$ number by the decay of these superheavy fermions, we do not discuss this
 possibility in more detail. We may return to this subject in future\footnote{
If the superheavy fields are vector, these operators must be decomposed as
two parts which include fermion and anti-fermion. Such decomposition is possible
for the operator
${\bf 10}\cdot{\bf 10}\cdot {\bf \bar 5}^\dagger\cdot{\bf\bar 5}^\dagger\cdot{\bf 5}^\dagger_s$. Actually, the vector bosons which belong to ${\bf 10}$ of $SU(5)$ can
induce the operator. }.

In this section, we have decomposed the dimension seven $B-L$ violating operators into
two parts. By this decomposition, we can address the generation of the $B-L$
number by the decay of the intermediate superheavy fields. 
Though some of the operators obtained by the decomposition have still higher dimension than four, we do not decompose them further because we do not need
the origin of the operators to discuss $B-L$ generation.


\section{$B - L$ Number Generation in the Early Universe}
In this section, we study the $B-L$ generation by the decay of the superheavy scalar
fields. 

First, let us fix the particle contents.  As discussed in the previous section, some additional fields are needed, and we introduce bosons denoted as
$Q$, $U^c$, $D^c$, $E^c$, $L$ whose charges are the same as the Standard Model fermions, $q$, $u_R^c$, $d_R^c$, $e_R^c$, $l$, respectively.

Next, we write down all the dimension four and five interactions which include only one superheavy scalar
as
\begin{itemize}
 \item dim. 4 :\\[1mm]
  $\quad d_R^clQ \: , \: d_R^cd_R^cU^c \: , \: llE^c \: , \: qlD^c \: , \:
   u_Rd_RD^c \: , \: q^{\dagger}q^{\dagger}D^c \: , \: u_R^{c\dagger}e_R^{c\dagger}D^c \: ,$\\[1mm]
  $\quad qd_R^cL \: , \: e_R^clL \: , \: q^{\dagger}u_R^{c\dagger}L \: , \:(h.c.)$,
 \item dim. 5 :\\[1mm]
  $\quad d_R^{c\dagger}d_R^{c\dagger}h_DQ \: , \: qqh_D^{\dagger}Q \: , \: u_R^ce_R^ch_D^{\dagger}Q \: , \:
   q^{\dagger}l^{\dagger}h_D^{\dagger}Q \: , \: u_R^{c\dagger}d_R^{c\dagger}h_D^{\dagger}Q \: ,$\\[1mm]
  $\quad d_R^{c\dagger}l^{\dagger}h_DU^c \: , \: qe_R^ch_D^{\dagger}U^c \: , \:
   u_R^{c\dagger}l^{\dagger}h_D^{\dagger}U^c \: , \:
   qu_R^ch_D^{\dagger}E^c \: , \: q^{\dagger}d_R^{c\dagger}h_D^{\dagger}E^c$\\[1mm]
  $\quad e_R^{c\dagger}l^{\dagger}h_D^{\dagger}E^c \: , \: d_R^{c\dagger}l^{\dagger}h_D^{\dagger}D^c \: , \:
  l^{\dagger}l^{\dagger}h_D^{\dagger}L \: , \: (h.c.)$,
\end{itemize}
where we omited the indices of spinor, gauge of $SU(2)$ and $SU(3)$ for simplicity.
\begin{table}[t]
 \begin{center}
  \begin{tabular}{|c||c|c|c|c|c|} \hline
   Interaction & $Q$ & $U^c$ & $E^c$ & $D^c$ & $L$\\ \hline \hline
   dim. 4 & $+4/3$ & $+2/3$ & $\:\:+2\:\:$ & $+2/3$ & $0$\\
   dim. 5 & $-2/3$ & $-4/3$ & $0$ & $-4/3$ & $\:\:-2\:\:$\\ \hline
  \end{tabular}  
  \caption{The generated $B - L$ number by decay of additional particles.}
  \label{table:B-L_number}
 \end{center}
\end{table}
It is interesting that the $B-L$ number of the final states by the decay of $Q$, $U^c$, $E^c$, $D^c$, $L$ 
is a fixed value for each superheavy fields and each dimension of the interactions as 
in Table \ref{table:B-L_number}.  Since 
the $B - L$ number of the final states induced by the dimension four interactions
is different from that by dimension five interactions, the decay can produce non-vanishing $B-L$ number.

For the estimation of the $B - L$ number, it is useful to calculate
the mean net $B - L$ number $\epsilon$;
\begin{equation}
 \epsilon_i = \sum_f x_{i \rightarrow f} \left[ r( i \rightarrow f ) - r( \bar{i} \rightarrow \bar{f} ) \right],
\end{equation}
where $i$ is the initial decay particle with mass of $m_i$, which contains $Q$, $U^c$, $E^c$, $D^c$, $L$.  $f$ means
the decay modes from the decay of $i$.  $x_{i \rightarrow f}$ is $B - L$ number within the decay modes $f$.  $r$ is the 
branching ratio; $\bar{i}$ or $\bar{f}$ means $CP$ transformated state, i.e., anti-particles.  $\epsilon_i$ means
generated $B - L$ number for the decay of two particles $i$ and $\bar{i}$. 
Therefore, we can obtain the $B-L$ number density $n_{B-L}$ from the number density of the $i$ particle $n_i$ and
the $\epsilon_i$ parameter as $n_{B-L}\sim \epsilon_in_i$. 
After the sphaleron process, the $B$ number density $n_B$
is obtained as 
\begin{equation}
n_B\sim 0.35 n_{B-L}\sim 0.35\epsilon_in_i. 
\end{equation}
Therefore, in order to obtain the $B$ number in a comoving frame $B\equiv\frac{n_B}{s}$, 
we have to know $\epsilon_i$ and $n_i$.

 For the calculation of $\epsilon_i$, we denote
couplings as follows;
\begin{center}
 \includegraphics[scale=0.75]{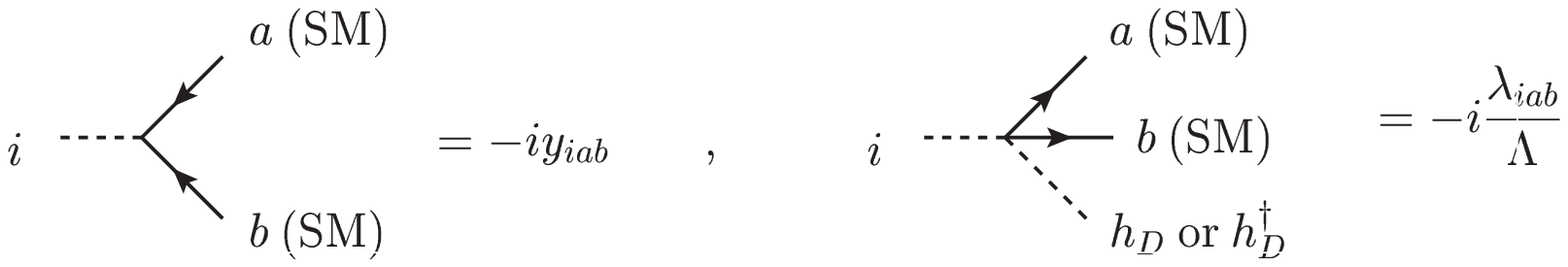} \label{couplings}
\end{center}
where $y$ and $\lambda$ are dimensionless couplings, and $\Lambda$ is the scale of the higher dimensional
interactions.
\begin{figure}[t]
 \begin{center}
  \includegraphics[scale=0.7]{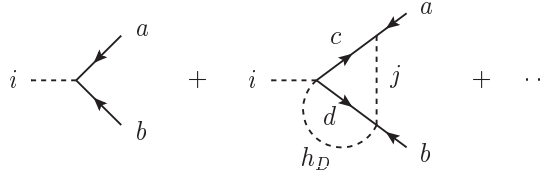}
  \caption{The Feynman diagrams for calculating $\epsilon_i$. }
  \label{epsilon}
 \end{center}
\end{figure}
By calculating the Feynman diagrams in Fig. \ref{epsilon}, we obtain $\epsilon_i$ as
\begin{equation}
 \epsilon_i = \frac{2}{(8\pi)^3} \frac{m_i^2}{\Lambda^2} \frac{m_i}{16\pi \Gamma_i}
  \sum_{j,a, \dots ,d} \textrm{Im} 
  \left( \lambda_{jab} \lambda_{icb}^{\dagger} y_{jcd}^{\dagger} y_{ida} \right) 
   \cdot f \left( m_j^2/m_i^2 \right).  \label{eq:epsilon}
\end{equation}
(See Appendix for the detail calculation.  As an example if we take $i=Q$ and $j=U^c$, the summation becomes
$\lambda_{Uld} \lambda_{Qdd}^{\dagger} y_{Udd}^{\dagger} y_{Qdl}+
\lambda_{Ulu} \lambda_{Qdu}^{\dagger} y_{Udd}^{\dagger} y_{Qdl}$. We can also take $j = E^c, D^c, L$ for $i=Q$.
Of course we can take $U^c$, $E^c$, $D^c$, $L$ as $i$. Once we fix the concrete fields as $i$ and $j$, the factor due 
to the number of freedom in the loop  is appearing. We just ignore it in the following for simplicity.)
$\Gamma_i$ is the total decay width;
\begin{equation}
 \Gamma_i = \frac{m_i}{16\pi} \sum_{a,b} 
 \left( y_{iab}^{\dagger} y_{iab} \right)
  + \frac{m_i}{3(8\pi)^3} \frac{m_i^2}{\Lambda^2} \sum_{a,b} 
  \left( \lambda_{iab}^{\dagger} \lambda_{iab} \right),
 \label{eq:G}
\end{equation}
where the first term is the contribution from the two body decay, and the second term is from the 
three body decay.
Function $f$ in (\ref{eq:epsilon}) is the loop function as follows;
\begin{eqnarray}
 f(\alpha) & \equiv & 
  1+2\alpha \left[ 1-(1+\alpha) \ln{\left(1+1/\alpha \right)} \right] \\
  & \sim & \left\{
   \begin{array}{lr}
    1 + \mathcal{O}(\alpha) & (\alpha \siml 1)\\[1mm]
    \frac{1}{3\alpha} + \mathcal{O}(1/\alpha^2) & (\alpha \simg 1)
   \end{array}
    \right. .
    \label{eq:loop_function}
\end{eqnarray}
If $m_i\sim m_j$ but $m_i<m_j$, then the function $f$ is rougly $\mathcal{O}(0.1)$.

Supposing that 
the only one coupling dominates the others for each $i$ particle
and for each dimensional operator. Namely, there are four couplings, $y_i$, $y_j$, $\lambda_i$, and $\lambda_j$.
Then, the eqs. (\ref{eq:epsilon}) and (\ref{eq:G}) can be rewritten as
\begin{eqnarray}
 \epsilon_i& = &\frac{2}{(8\pi)^3} \frac{m_i^2}{\Lambda^2} \frac{m_i}{16\pi \Gamma_i}
 \textrm{Im} 
  \left( \lambda_j\lambda_i^{\dagger} y_j^{\dagger} y_i \right) 
   \cdot f \left( m_j^2/m_i^2 \right), \label{eq:epsilonS} \\
  \Gamma_i& =& \frac{m_i}{16\pi}
  |y_i|^2
  + \frac{m_i}{3(8\pi)^3} \frac{m_i^2}{\Lambda^2} 
  |\lambda_i|^2. \label{eq:GS}
\end{eqnarray} 
Moreover, 
if we take $y\equiv|y_i|\sim |y_j|$ and 
$\lambda\equiv|\lambda_i|\sim |\lambda_j|$ and the branching ratio of two body decay is comparable
to that of three body decay, i.e., $y\sim m_i\lambda/(4\sqrt{6}\pi\Lambda)$,
then we can obtain simpler equations as 
\begin{eqnarray}
 \Gamma_i &\sim& 2\frac{m_i}{16\pi}y^2. \label{eq:G_appro1} \\
  \epsilon_i &\sim& \frac{3}{16\pi}y^2f\sin\delta,\label{eq:e_appro1}
\end{eqnarray}
where
$\sin\delta\equiv\textrm{Im} \left( \lambda_j\lambda_i^{\dagger} y_j^{\dagger} y_i \right)/(y^2\lambda^2)$.

Next, let us estimate the abundance of the $i$ particle, $n_i$, and the Baryon number in a comoving frame $B$
in the following two cases.
In the first case, the particle $i$ is thermally produced, the freeze out occurs 
when the particle $i$ is still relativistic, and no entropy is produced by the decay (case A).
 (Therefore, we assume that the
reheating temperature due to the inflation is larger than the mass of the superheavy particle $i$.
We discuss whether the particle $i$ is still relativistic or not at the freeze out
in section 4.)
Then, $Y_i\equiv \frac{n_i}{s}$ is given by $Y_i\sim 0.278\frac{g_{eff}}{g_{*S}}$, where the entropy
density $s$ is obtained as $s=\frac{2\pi^2}{45}g_{*S}T^3$ and $g_{eff}$ is the number of freedom of $i$.
Therefore, we can obtain
\begin{equation}
B\sim 0.1 \frac{g_{eff}\epsilon_i}{g_{*S}}\sim 3.5\times 10^{-5}y^2,
\end{equation}
where in the last similarity we use eq. (\ref{eq:e_appro1}), $f\sin\delta\sim 0.1$, $g_{*S}\sim 100$, and
$g_{eff}\sim 6$ for $i=Q$. Therefore, 
\begin{equation}
y\sim 2\times 10^{-3} \label{cond1}
\end{equation} is required to obtain $B\sim 10^{-10}$.
An additional condition $\Gamma_i>\langle \sigma v_i\rangle n_i$ is required so that the estimation 
$n_{B-L}\sim \epsilon_in_i$ is valid,
where $\sigma$ and $v_i$ are the cross section of the annihilation
 process and the velocity of the particle $i$, respectively.
If $\langle\sigma v_i\rangle\sim 0.01T_D/m_i^3$, this condition is roughly rewritten as
\begin{equation}
\left(\frac{m_i}{M_{pl}}\right)^2>10^{-6}y^2.
\label{cond2}
\end{equation}
Here, the decay temperature $T_D$ is defined by the temperature of the universe when 
the age of the universe is around the lifetime of the particle $i$, which
 is given by
$T_D\sim\sqrt{\Gamma_iM_{pl}/1.66\sqrt{g_*}}$, where $M_{pl}$ is the Planck mass as 
$M_{pl} = 1.22 \times 10^{19} \: \rm{GeV}$. 
From the eq. (\ref{cond1}), the inequality (\ref{cond2}) is rewritten as
\begin{equation}
m_i>2\times 10^{-6} M_{pl}\sim 2\times 10^{13} \: \rm{GeV}.
\end{equation}
When $m_i\sim 10^{14}(10^{16})$ GeV, 
$\Gamma_i\sim 10^{-7} m_i\sim 10^7 (10^9)$ GeV, and therefore, 
$T_D\sim 3\times 10^{12}(3\times 10^{13})$ GeV.
The higher dimensional coupling is given as
$\lambda/\Lambda\sim 4\sqrt{6}\pi y/m_i\sim 5\times 10^{-16} (5\times 10^{-18}) ({\rm GeV})^{-1}$.

As the second case (case B),  
we consider the situation in which the density of $i$ and $\bar i$ fields
dominates the density of the universe. Generically, thermal abundance of the heavy particle
with long lifetime becomes large and sometimes dominates the energy density of the universe.
Then, 
\begin{equation}
\rho=\rho_i+\rho_{\bar i}=2m_in_i=\frac{\pi^2}{30}g_*T_R^4, 
\label{TR}
\end{equation}
where $\rho_i$, $\rho_{\bar i}$, $n_i$, $g_*$, and $T_R$ are
the energy density of $i$ field, that of $\bar i$ field, the number density of $i$ field, 
the total number of effectively massless degrees of freedom, and the temperature after the
$i$ and $\bar i$ field decay, respectively. 
The $B-L$ number in a comoving volume is given as
\begin{equation}
\frac{n_{B-L}}{s}=\frac{3}{8}\frac{g_*}{g_{*S}}\frac{T_R}{m_i}\epsilon_i.
\end{equation}
After the sphaleron process, the $B$ number in a comoving volume is given as
\begin{equation}
B\equiv\frac{n_B}{s}\sim 0.35\frac{n_{B-L}}{s}\sim \frac{1}{8}\frac{T_R}{m_i}\epsilon_i,
\end{equation}
where we took $g_* \sim g_{*S}$.
 Therefore, the Baryon number is given by
\begin{equation}
B=\frac{3}{128\pi}\frac{y^2T_R}{m_i}f\sin\delta
= \frac{3y^3}{256\sqrt{3.32}\pi^{1.5}g_*^{1/4}}\sqrt{\frac{M_{pl}}{m_i}}f\sin\delta\sim
4 \times 10^{-5}\sqrt{\frac{M_{pl}}{m_i}}y^3,
\end{equation}
where 
the last similarity is given by taking
$f\sin\delta\sim 0.1$ and $g_*= \mathcal{O}(100)$. Roughly, if we take
\begin{equation}
y^3\sqrt{M_{pl}/m_i}\sim 3 \times 10^{-6}, 
\label{cond3}
\end{equation}
then we can obtain $B\sim 10^{-10}$.
The additional condition 
$\Gamma_i>\langle \sigma v_i\rangle n_i$ becomes 
\begin{equation}
\left(\frac{m_i}{M_{pl}}\right)^5>10^{-12}y^6
\label{cond4}
\end{equation}
by using eq. (\ref{TR}). 
From the eq. (\ref{cond3}), the additional condition (\ref{cond4}) is rewritten as
\begin{equation}
m_i>2\times 10^{-6} M_{pl}\sim 2\times 10^{13} \: \rm{GeV}.
\end{equation}
When $m_i\sim 10^{14}(10^{16})$ GeV, the eq. (\ref{cond3}) results in $y\sim 2(4)\times 10^{-3}$.
Then $\Gamma_i\sim 2\times 10^{-7}(7\times 10^{-6}) m_i\sim 2\times 10^7 (7\times 10^{9})$ GeV, and therefore, 
$T_R\sim 3\times 10^{12}(7\times 10^{13})$ GeV.
The higher dimensional coupling is given as
$\lambda/\Lambda\sim 4\sqrt{6}\pi y/m_i\sim 6\times 10^{-16} (10^{-17}) ({\rm GeV})^{-1}$.

In this section, we have shown that the Baryon asymmetry in the universe can be explained
by the $B-L$ production by the decay of some superheavy particle which can exist 
in some GUT models.
.

\section{Discussion and Summary}
The initial density of the superheavy fields may be produced non-thermally like the 
preheating\cite{preheating} and dominate the density of the universe. But here we consider
the thermal abundance of the superheavy fields.
If we take $\langle \sigma v_i\rangle\sim 0.01\frac{T}{(m_i^2+T^2)^{\frac{3}{2}}}$, $g_*=g_{*S}=106.75$,
and $g_{eff}=1$, then the numerical calculation shows that when the mass $m_i$ is larger than $10^{14}$
GeV the number density of the $i$ particle behaves like hot relics as $Y_i\sim 2\times 10^{-3}$
as in Table \ref{table:thermal}.
\begin{table}[t]
 \begin{center}
  \begin{tabular}{|c||c|c|c|c|c|c|c|} \hline
   $m_i$(GeV) & $10^{11}$ & $10^{12}$ & $10^{13}$ & $10^{14}$ & $10^{15}$ & $10^{16}$ & Hot relics \\ \hline \hline
   $Y_i$($\times 10^{-3}$) & $ 0.019$ & $0.10$ & $0.46$ & $1.4$ 
   & $2.0$ & $2.1$ & $2.2$ \\ \hline
  \end{tabular}  
  \caption{Thermal abundance $Y_i$ with $y=0$.}
  \label{table:thermal}
 \end{center}
\end{table}
In the numerical calculation, we used Boltzmann equations with Maxwell-Bolzmann approximation for the distribution function.
Therefore, for this mass range, the calculation in case A is reasonable. But if 
$\frac{\rho_i+\rho_{\bar i}}{\rho_R}=\frac{8}{3}\frac{g_{eff}g_{*S}}{g_*}\frac{m_i}{T_R}Y_i>1$, 
then the calculation in case B is preferable because of the entropy production due to the decay
of the particle $i$.
Therefore, we conclude that thermal abundance of the superheavy fields is sufficient to explain
the Baryon assymmetry in the universe. One of the point is that since the particles
are superheavy, the Hubble expansion rate becomes so high that even gauge interactions
can be out of equilibrium and not affect the generation of asymmetry. This is quite different
from the usual leptogenesis.

If one of the right-handed neutrino masses is smaller than the decay temperature
and $10^{12}$ GeV at which the sphaleron process becomes thermalized, then the produced
$B-L$ number is washed out by the equilibrium of the $B-L$ violating neutrino process and the
shaleron process\cite{washout}. Therefore, the mass of the right-handed neutrinos must be larger than the
decay temperature in order to obtain the Baryon assymmetry by this mechanism
if the right-handed neutrino is lighter than $10^{12}$ GeV.

The $B-L$ violating dimension 7 interactions via the superheavy fields, whose couplings are
$y\lambda/(m_i^2\Lambda)$,  can induce the
instability of the nucleon. However, the contribution is negligible because the effective
dimension 6 couplings become very small as
$y\lambda\langle h_D\rangle/(m_1^2\Lambda)\ll 1/M_{pl}^2$. 

In this paper, we do not introduce the supersymmetry(SUSY), but the extension to the SUSY models
is straightforward. In some SUSY GUT models\cite{anomalousGUT}, there may be superheavy fields,
$\bf 10+\bar {10}$, $\bf 24$, and $\bf 5+\bar 5$ of $SU(5)$, some of which may produce the
Baryon assymmetry in the universe, though the serious gravitino problem must be taken
into account\cite{gravitino}.

In this paper, we have studied the possibility that the decay of the superheavy particles,
which may be induced in grand unified theories as extra fields, produces the non-vanishing 
$B-L$ number, which converts to the Baryon assymmetry by the shaleron process.
We have shown that if the mass of the superheavy field is larger than $10^{13-14}$ GeV, 
the Baryon assymmetry in the universe can be explained by the decay of the superheavy field
with appropriate couplings.

\section*{Acknowledgments}
We thank M. Tanabashi and T. Yamashita for valuable comments.
N.M. is supported in part by Grants-in-Aid for Scientific Research from
MEXT of Japan.
This work was partially supported by the Grand-in-Aid for Nagoya
University Global COE Program,
``Quest for Fundamental Principles in the Universe:
from Particles to the Solar System and the Cosmos'',
from the MEXT of Japan.

\appendix

\section{The Calculation of the Mean Net $B - L$ Number}
In this appendix, we will calculate the mean net $B- L$ number $\epsilon_i$ defined by
\begin{equation}
 \epsilon_i = \sum_f x_{i \rightarrow f} ( r_{i \rightarrow f} - r_{\bar{i} \rightarrow \bar{f}}).
 \label{eq:mean_net_number_app}
\end{equation}

Firstly let us simplify (\ref{eq:mean_net_number_app}).  The sum of the branching ratios of a group of decay modes $g$ which have the same $B-L$ charges $x_{i\rightarrow g}=const$ 
can be written
\begin{equation}
 \sum_g r_{i \rightarrow g} = 1 - \sum_{f \neq g} r_{i \rightarrow f}.
\end{equation}
Therefore, (\ref{eq:mean_net_number_app}) is transformed to
\begin{eqnarray}
 \epsilon_i &=& x_{i \rightarrow g} \sum_g ( r_{i \rightarrow g} - r_{\bar{i} \rightarrow \bar{g}} )
  + \sum_{f \neq g} x_{i \rightarrow f} ( r_{i \rightarrow f} - r_{\bar{i} \rightarrow \bar{f}} ) \\
  &=& - x_{i \rightarrow g} \sum_{f \neq g} ( r_{i \rightarrow f} - r_{\bar{i} \rightarrow \bar{f}} )
  + \sum_{f \neq g} x_{i \rightarrow f} ( r_{i \rightarrow f} - r_{\bar{i} \rightarrow \bar{f}} ) \\
  &=& \sum_{f \neq g} (x_{i \rightarrow f}- x_{i \rightarrow g})( r_{i \rightarrow f} - r_{\bar{i} \rightarrow \bar{f}} ).
  \label{eq:epsilon_simplfy}
\end{eqnarray}
Therefore, in calculating $\epsilon_i$, we do not have to calculate all the branching ratios.

In our case, modes $f$ run two and three body decays which are induced by dim.4 and 5 interactions respectively. Here,
we choose the three body decays as modes $g$\footnote{Of course, it is the same results if you choose the two body decays
as modes $g$.}.  Since $x_{i \rightarrow 2bd.} - x_{i \rightarrow 3bd.} = 2$ for all species $i$ as is shown in Table
\ref{table:B-L_number}, we obtain
\begin{eqnarray}
 \epsilon_i &=& 2 \sum_{f=2bd.} ( r_{i \rightarrow f} - r_{\bar{i} \rightarrow \bar{f}} )\\
 &=& 2 \sum_{f=2bd.} ( \Gamma_{i \rightarrow f} - \Gamma_{\bar{i} \rightarrow \bar{f}} ) / \Gamma_i
\end{eqnarray}
from (\ref{eq:epsilon_simplfy}), where $\Gamma_{i \rightarrow f}$ is the partial decay width and $\Gamma_i$ is the total
decay width defined by
\begin{equation}
 \Gamma_i \equiv \sum_f \Gamma_{i \rightarrow f} = \sum_f \Gamma_{\bar{i} \rightarrow \bar{f}} \label{eq:differential_width}
\end{equation}

Next, let us calculate difference of partial decay width $\Gamma_{i \rightarrow f} - \Gamma_{\bar{i} \rightarrow \bar{f}}$.
In case of two body decays, the width is given by
\begin{equation}
 \Gamma_{i \rightarrow f} = \frac{1}{16 \pi m_i} \left| \mathcal{M}_{i \rightarrow f} \right|^2, \label{eq:width}
\end{equation}
 where $\mathcal{M}$ is the amplitude. Here we assume that the decay products are massless.    Using the following Feynman rules
\begin{center}
 \includegraphics[scale=0.75]{couplings}
\end{center}
the amplitude can be calculate as follows;
\begin{eqnarray}
 \left| \mathcal{M}_{i \rightarrow ab} \right|^2 &=&
  \begin{array}{|c|} \includegraphics[scale=0.75]{amplitude} \end{array}^{\:2}\\
 &=& 2(p_a \cdot p_b) \left| y_{iab} + \sum_{j,c,d} y_{jac}
  \frac{\lambda_{icd}}{\Lambda} \frac{\lambda_{jdb}^{\dagger}}{\Lambda} \mathcal{F}(m_i,m_j) + \cdots \right|^2\\
 &=& m_i^2 \left[ \left| y_{iab} \right|^2 + 2 \textrm{Re} \sum_{j,c,d} y_{iab}^{\dagger} y_{jac}
  \frac{\lambda_{icd}}{\Lambda} \frac{\lambda_{jdb}^{\dagger}}{\Lambda} \mathcal{F}(m_i,m_j) + \cdots \right],
 \label{eq:amplitude}
\end{eqnarray}
where $\mathcal{F}$ is loop function.  On the other hands, the amplitude for the anti-particle
can be obtained by taking hermite conjugated couplings from the amplitude for the particle;
\begin{eqnarray}
 \left| \mathcal{M}_{\bar{i} \rightarrow \bar{a}\bar{b}} \right|^2 
 &=& 2(p_a \cdot p_b) \left| y_{iab}^{\dagger} + \sum_{j,c,d} y_{jac}^{\dagger}
  \frac{\lambda_{icd}^{\dagger}}{\Lambda} \frac{\lambda_{jdb}}{\Lambda} \mathcal{F}(m_i,m_j) + \cdots \right|^2\\
 &=& m_i^2 \left[ \left| y_{iab} \right|^2 + 2 \textrm{Re} \sum_{j,c,d} y_{iab}^{\dagger} y_{jac}
  \frac{\lambda_{icd}}{\Lambda} \frac{\lambda_{jdb}^{\dagger}}{\Lambda} \mathcal{F}^{\dagger}(m_i,m_j) + \cdots \right],
 \label{eq:anti_amplitude}
\end{eqnarray}
Using (\ref{eq:width}), (\ref{eq:amplitude}) and (\ref{eq:anti_amplitude}), the difference of partial decay width can be
written as
\begin{eqnarray}
 \Gamma_{i \rightarrow ab} - \Gamma_{\bar{i} \rightarrow \bar{a}\bar{b}} &=&
  \frac{1}{1 6\pi m_i} \left( \left|\mathcal{M}_{i \rightarrow ab} \right|^2
   - \left|\mathcal{M}_{\bar{i} \rightarrow \bar{a}\bar{b}} \right|^2 \right)\\
 &=& -\frac{m_i}{4\pi} \sum_{j,c,d} \textrm{Im} \left(
   y_{iab}^{\dagger} y_{jac} \frac{\lambda_{icd}}{\Lambda} \frac{\lambda_{jdb}^{\dagger}}{\Lambda} \right)
    \textrm{Im} \mathcal{F}(m_i,m_j) + \cdots \label{eq:differential_width2}
\end{eqnarray}
Here, $\textrm{Im}\mathcal{F}$ is given by
\begin{equation}
 \textrm{Im}\mathcal{F} = - \frac{m_i^2}{2(8\pi)^3} f \left( m_j^2/m_i^2 \right),\label{eq:ImF}
\end{equation}
where $f$ is the function defined by (\ref{eq:loop_function}). Note that the function $\mathcal{F}$
is diverging but $\textrm{Im}\mathcal{F}$ becomes finite. This is because the imaginary part can 
be estimated just by tree diagrams if Cutkosky rules are applied.

Finally, we can obtain the mean net $B - L$ number using (\ref{eq:epsilon_simplfy}), (\ref{eq:differential_width2}) and
(\ref{eq:ImF});
\begin{eqnarray}
 \epsilon_i &=& \frac{2}{(8\pi)^3} \frac{m_i^2}{\Lambda^2} \frac{m_i}{16 \pi \Gamma_i} \sum_{j,a,\cdots,d}
  \textrm{Im} \left( y_{iab}^{\dagger} y_{jac} \lambda_{icd} \lambda_{jdb}^{\dagger} \right)
   f \left( m_j^2/m_i^2 \right).
\end{eqnarray}

\end{document}